\documentstyle[12pt]{article}

\newcommand{\eq}[1]{Eq.~(\ref{#1})}

\newcommand{\nl}{\nonumber \\}

\def\beq{\begin{equation}}
\def\eeq{\end{equation}}
\def\beqa{\begin{eqnarray}}
\def\eeqa{\end{eqnarray}}

\newcommand{\sect}[1]{\setcounter{equation}{0}\section{#1}}

\newcommand{\EQ}{\begin{equation}}
\newcommand{\EN}{\end{equation}}
\newcommand{\bea}{\begin{eqnarray}}
\newcommand{\ena}{\end{eqnarray}}

\renewcommand{\a}{\alpha}

\renewcommand{\thefootnote}{\fnsymbol{footnote}}

\begin{document}
\begin{titlepage}
\rightline{DFTT 41/96}
\rightline{\hfill July 1996}

\vskip 1.2cm

\centerline{\Large \bf Two-loop scalar diagrams }
\centerline{\Large \bf from string theory}

\vskip 1.2cm

\centerline{\bf Paolo Di Vecchia\footnote{e-mail:
DIVECCHIA@nbivms.nbi.dk}
 and Lorenzo Magnea\footnote{On leave from Universit\`a di Torino, Italy}}
\centerline{\sl NORDITA}
\centerline{\sl Blegdamsvej 17, DK-2100 Copenhagen \O, Denmark}

\vskip .2cm

\centerline{\bf Alberto Lerda\footnote{II Facolt\`a di Scienze
M.F.N., Universit\`a di Torino
(sede di Alessandria), Italy}}
\centerline{\sl Dipartimento di Scienze e Tecnologie Avanzate and}
\centerline{\sl Dipartimento di Fisica Teorica, Universit\`a di Torino}
\centerline{\sl Via P.Giuria 1, I-10125 Torino, Italy}
\centerline{\sl and I.N.F.N., Sezione di Torino}

\vskip .2cm

\centerline{\bf Raffaele Marotta\footnote{Della Riccia fellow}}
\centerline{\sl Dipartimento di Scienze Fisiche, Universit\`a di Napoli}
\centerline{\sl Mostra D'Oltremare, Pad. 19, I-80125 Napoli, Italy}

\vskip .2cm

\centerline{\bf Rodolfo Russo}
\centerline{\sl Dipartimento di Fisica, Politecnico di Torino}
\centerline{\sl Corso Duca degli Abruzzi 24, I-10129 Torino, Italy}
\centerline{\sl and I.N.F.N., Sezione di Torino}
\vskip 1cm

\begin{abstract}

We show how to obtain correctly normalized expressions for the Feynman 
diagrams of $\Phi^3$ theory with an internal $U(N)$ symmetry group,
starting from tachyon amplitudes of the open bosonic string, and  
suitably performing the zero--slope limit by giving an                    
arbitrary mass $m$ to the tachyon. In particular we 
present explicit results for the two--loop amplitudes of    
$\Phi^3$ theory, in preparation for the more interesting case of the multiloop
amplitudes of non--abelian gauge theories.  

\end{abstract}

\end{titlepage}

\newpage
\renewcommand{\thefootnote}{\arabic{footnote}}
\setcounter{footnote}{0}
\setcounter{page}{1}

\sect{Introduction}
\label{intro}
\vskip 0.5cm

It has been stressed by several people that string theory can
not only be used as a laboratory for a realistic unifying theory of all
interactions, but can also provide an efficient way of
computing  perturbative field theory amplitudes in the limit
of vanishing Regge slope $\a ' \rightarrow 0$.
This approach, pioneered for gluon amplitudes
at tree level by the authors of
Ref.~\cite{manpar}, was extensively applied to one--loop
amplitudes~\cite{berkos}, leading to a set
of simple string--inspired ``Feynman'' rules.
Subsequently, the first steps have been taken towards
a generalization to multiloops~\cite{kaj}, mainly using the technology
of the multiloop operator formalism for string
amplitudes~\cite{copgroup}.
These string--derived methods have also inspired the development
of new techniques
in field theory, which make use of world--line path integrals.
This latter approach was studied at one loop in Ref.~\cite{strass},
and partially extended to higher
loops in Ref.~\cite{scsc}. Recently, a connection between the world--line
Green functions used in field theory and the multiloop
Green function of string theory
has been established in Ref.~\cite{kajsat}.

In two previous papers, we performed the field theory limit in the
amplitudes with two~\cite{letter}, three and
four gluons~\cite{old1} in the open bosonic string,
and extracted the corresponding
renormalization constants after a suitable off--shell extension.
These results were obtained by discarding by hand the contribution
of the tachyons, and isolating the contribution of the gluons (the
higher string states are always negligible when $\a' \rightarrow 0$).
In Refs.~\cite{letter,old1} we also stressed that, in order to
compute gluon amplitudes in field theory, one is not forced to start
from a consistent string model; indeed, it was practical to use the
simplest  model
containing Yang-Mills theory when $\a' \rightarrow 0$,
{\it i.e.} the open bosonic string.

In this paper we show how these results can be extended to higher loops.
As a first step, to avoid the computational difficulties associated with
multiloop Yang--Mills amplitudes, which are, however, inessential for the
understanding of the field theory limit, we focus on scalar particles
instead of gluons. This can be achieved by considering a
slightly different zero--slope limit of the bosonic string,
in which only the lowest tachyonic state, with a mass $m$ satisfying
$\a' m^2 = -1$, is kept. To avoid the inconsistencies associated with
tachyon  propagation, one must carefully modify the measure of integration
by replacing the tachyon poles with exponentials of the
mass of the lowest state.
In the case of tree and one--loop
diagrams, this procedure is equivalent to taking the
zero--slope limit of an old
pre--string dual model characterized by an arbitrary
value of the intercept  of the Regge trajectory
$a$ \footnote{Note that
$a$ is related to the mass of the lowest scalar state of
the theory by $a + \a' m^2 =0$.}. This model was recognized to be
inconsistent, because
of the presence of ghost states unless $a=1$, in which case, however,
the lowest scalar particle
becomes a tachyon. The tree and one--loop diagrams of
this pre--string dual
model  were shown to lead to the Feynman diagrams of $\Phi^3$ theory
by Scherk~\cite{Scherk}.

In this paper we explicitly show, up to two loops, that, by performing a
zero--slope limit in which we keep only a suitably reinterpreted
tachyon state,
one correctly reproduces the Feynman diagrams of the $\Phi^3$ theory, with
the inclusion  of an internal symmetry group $U(N)$, and with the right
overall normalization. In the process we also develop a precise understanding
of
which corners of the multiloop moduli space contribute to the field theory
limit, and propose an algorithm that allows to obtain
all diagrams of the $\Phi^3$ theory in
the Schwinger parametrization (after integration of all loop momenta) from the
general string amplitudes.
We believe that this algorithm will be easily generalizable to Yang--Mills
theory, where we can expect extra difficulties due to the complicated
numerator structure of the integrands, but the
nature of the field theory limit
should be unchanged.

\sect{Scalar amplitudes in string theory}
\label{tach}

In Ref.~\cite{old1} we have given the correctly normalized $h$--loop
$M$--gluon
amplitude in the open bosonic string. That formula can be
immediately adapted to the case of scalars. The planar $h$--loop
scattering amplitude of $M$ scalars with momentum $p_1,\ldots,p_M$, is
\bea
A^{(h)}_M (p_1,\ldots,p_M) & = & N^h\,{\rm Tr}(\lambda^{a_1}
\cdots \lambda^{a_M})~
C_h\,\left[2 g_s \left( 2 \a' \right)^{(d-2)/4} \right]^M    \nl
& \times & \int [dm]^M_h\,
\prod_{i<j} \left[{{\exp\left({\cal G}^{(h)}(z_i,z_j)\right)}
\over{\sqrt{V'_i(0)\,V'_j(0)}}}\right]^{2\a' p_i\cdot p_j}~~,
\label{hmastac}
\ena
where $N^h\,{\rm Tr}(\lambda^{a_1}
\cdots \lambda^{a_M})$ is the appropriate $U(N)$ Chan-Paton factor
\footnote{We adopt the same normalizations and conventions of
ref.~\cite{old1}, in particular ${\rm Tr}(\lambda^a \,\lambda^b)=
\frac{1}{2}\delta^{ab}$.}, $g_s$ is the dimensionless string coupling constant,
${\cal C}_h$ is a normalization factor given by
\EQ
C_h = {1\over{(2\pi)^{dh}}}~g_s^{2h-2}{1\over{(2\a')^{d/2}}}~~,
\label{vertnorm}
\EN
and ${\cal G}^{(h)}$ is the $h$--loop bosonic Green function
\beq
{\cal G}^{(h)}(z_i,z_j) = \log E^{(h)}(z_i,z_j) - {1\over 2} \int_{z_i}^{z_j}
\omega^\mu \, \left(2\pi {\rm Im}\tau_{\mu\nu}\right)^{-1}
\int_{z_i}^{z_j} \omega^\nu~~~,
\label{hgreen}
\eeq
with $E^{(h)}(z_i,z_j)$ being the prime form,
$\omega^\mu$ ($\mu=1,\ldots, h$) the abelian
differentials and $\tau_{\mu\nu}$ the period matrix.
All these objects, as well as the measure on moduli space $[dm]^M_h$, can
be explicitly written in the Schottky parametrization of the Riemann
surface, and their expressions for arbitrary $h$ can be found for example
in Ref.~\cite{scho}. Here we give only the
expression for the measure, which for scalar amplitudes
differs slightly from
the one of Ref.~\cite{old1}. It is given by
\beqa
[dm]^M_h & = & \frac{1}{dV_{abc}} \prod_{i=1}^M \frac{dz_i}{ V_{i} ' (0)
 }
\prod_{\mu=1}^{h} \left[ \frac{dk_\mu \,d \xi_\mu \,d \eta_\mu}{k_\mu^2
\,(\xi_\mu - \eta_\mu)^2} ( 1- k_\mu )^2 \right]   \label{hmeasure} \\
& \times & \left[\det \left( - i \tau_{\mu \nu} \right) \right]^{-d/2}
{}~\prod_{\alpha}\,' \left[ \prod_{n=1}^{\infty} ( 1 - k_{\alpha}^{n})^{-d}
\prod_{n=2}^{\infty} ( 1 - k_{\alpha}^{n})^{2} \right]~~~.   \nonumber
\eeqa
where $k_{\mu}$ are the multipliers, $\xi_{\mu}$ and $\eta_{\mu}$ are the
fixed points of the generators of the Schottky group,
and $dV_{abc}$ is the
projective invariant volume element
\beq
dV_{abc} = \frac{d\rho_a~d\rho_b~d\rho_c}
{(\rho_a-\rho_b)~(\rho_a-\rho_c)~(\rho_b-\rho_c)}~~~,
\label{projvol}
\eeq
with $\rho_a$, $\rho_b$, $\rho_c$ being any three of the $M$
Koba-Nielsen variables $z_i$, or of the $2h$ fixed
points of the generators of the
Schottky group, which can be fixed at will. In the last line
of \eq{hmeasure}, the primed product
over $\alpha$ denotes the product over the primary
classes of elements of the
Schottky group, as defined in Ref.~\cite{scho}. Notice that
the measure (\ref{hmeasure}) contains an extra factor of
$\prod_{i=1}^M [V_{i}'(0)]^{-1}$, which originates
from the $N$-reggeon vertex, as explained in
Ref.~\cite{copgroup} \footnote{For gluons, this extra factor is
cancelled since
the amplitude is multilinear in every gluon polarization vector
$\varepsilon_i$ and each one of these is accompanied by
an extra $V_{i}'(0)$.}. Finally, like in any planar
open string amplitude, here too the Koba-Nielsen variables $z_i$
must be cyclically ordered
along one of the boundaries of the world--sheet.

Before proceeding, a few comments are in order.
First of all, \eq{hmastac} represents the
scattering amplitude of $M$ tachyons
if we require that the bosonic string be
consistent. In fact, in order not to have ghosts at the excited levels,
the intercept of the Regge trajectory, $a$, must be fixed to 1.
In this way, however, the mass squared of the scalar states
turns out to be negative: $\a'p_i^2=-\a'm^2=1$.
Secondly, the amplitude in \eq{hmastac} depends, in general,
on the projective transformations  $V_i (z)$, that
parametrize a choice of local coordinates
around the punctures $z_i$. However, by imposing the mass--shell
condition, $\a' p_{i}^{2} =1$, this dependence disappears, just as it
did for gluons.

As stressed in the introduction, our main purpose is to study
the zero--slope limit of \eq{hmastac}. Since in this limit
all excited string states (including possible ghost states)
become irrelevant, we can try to relax
the condition $a = 1$, and verify whether this leads to any
inconsistencies in the field theory limit. The analysis of string amplitudes
with an arbitrary value of the intercept of the Regge
trajectory was actually performed
long ago in the context of dual models, at least at tree
level and at one loop. Details on these amplitudes can be found,
for example, in the old review paper in Ref.~\cite{amaale}.
We will thus begin by borrowing some of the results of these models,
whose lowest state is a scalar particle with a mass $m$ such that
\beq
a + \a ' m^2 =0
\label{onshe}
\eeq
The on--shell tree--level $M$--point scalar amplitude is
\beqa
A^{(0)}_{M}(p_1,\ldots,p_M)
 & = & {\rm Tr}(\lambda^{a_1} \cdots \lambda^{a_M})\,  \frac{1}{
g_{s}^{2} (2 \a' )^{d/2}} \left[ 2 g_s
\left( 2 \alpha '\right)^{(d-2)/4} \right]^M
\nonumber \\
& \times & \int \frac{\prod\limits_{i=1}^{M} d z_i }{dV_{abc}} \prod_{i=1}^{M}
| z_{i-1} - z_i |^{a-1}
\prod_{i<j} \left( z_i - z_j \right)^{2 \alpha' p_i \cdot p_j }~~,
\label{tree}
\eeqa
which obviously reduces to the $M$-tachyon amplitude in the open bosonic
string if $a=1$.
In particular for $M=3$, we get
\beq
A^{(0)}_{3} = {\rm Tr}(\lambda^{a_1} \lambda^{a_2} \lambda^{a_3})\,
8 g_s \left( 2 \alpha '\right)^{(d-6)/4}~~.
\label{3tree}
\eeq
We observe that this same expression can be obtained by computing the
color--ordered
3--point function in the theory defined by the Lagrangian
\beq
L = {\rm Tr} \left[ \partial_{\mu} \Phi \partial^{\mu} \Phi + m^2 \Phi^2
- \frac{g}{3!} \Phi^3  \right]~~,
\label{Phi3lag}
\eeq
where $\Phi=\Phi^a\lambda^a$,
if the coupling constant $g$ and the string
coupling $g_s$ are related by
\beq
g = 16 g_s \left( 2 \alpha ' \right)^{(d-6)/4}~~.
\label{couplingide}
\eeq
Thus in the limit $\a '\to 0$, the amplitudes in \eq{tree} yield the
tree--level Feynman diagrams of the $\Phi^3$ theory (\ref{Phi3lag}).

We now consider the one--loop corrections.
The one--loop $M$--point scalar amplitude can also be found in
Ref.~\cite{amaale}.
We rewrite it here including the Chan-Paton factors,
the correct overall normalization, and choosing the fixed points
of the generator of the Schottky group $\xi= \infty$ and $\eta =0$,
together with $z_1 = 1$.
We find
\bea
A^{(1)}_M (p_1,\ldots,p_M) & = & N \, {\rm Tr}(\lambda^{a_1}
\cdots \lambda^{a_M}) ~
 \frac{1}{(4\pi)^{d/2}}\left(\frac{g}{8}\right)^M
( 2 \alpha ')^{M-d/2} \nl
& \times & \int_{0}^{1}  \frac{dk}{k^{a+1}} \,
\left[ - \frac{1}{2} \log k \right]^{-d/2}
\prod_{n=1}^\infty \left(1 - k^n  \right)^{-d} \nl
& \times & \int_{k}^{1} d z_{M} \int_{z_M}^{1}
d z_{M-1} \dots
\int_{z_3}^{1} d z_2~
\prod_{i=1}^{M} \frac{1}{\left[  V_{i} ' (0) \right]^{a}} \label{oneloop1} \\
& \times & \prod_{i<j} \left[{{\exp\left({\cal G}^{(h)}(z_i,z_j)\right)}
\over{\sqrt{ V'_i(0)\, V'_j(0)}}}\right]^{2\a' p_i\cdot p_j}
{}~\prod_{i=2}^{M+1} \left| z_{i-1} - z_i \right|^{a-1}~~,
\nonumber
\ena
where we have defined $z_{M+1} \equiv k$. Notice that \eq{oneloop1} is
derived by including only orbital degrees of freedom, and without cancelling
any unphysical state circulating in the loop (such states will
anyway be irrelevant in the field theory limit).
If one compares \eq{oneloop1} with the corresponding one--loop
expression that can be obtained from Eqs.~(\ref{hmastac}) and
(\ref{hmeasure}), one sees two discrepancies.
The first one is the absence, in the open string amplitude,
of the last term in \eq{oneloop1}.
The second one is the power of
$k^{- 1 - a}$ in the measure in \eq{oneloop1}, instead of $k^{-2}$,
Obviously, the two expressions agree if $a=1$, as expected.
Actually there is a third difference between the two
one--loop formulas, because the ghost contribution has been included
in one case and not in the other; however, for the scalar amplitudes, this
will irrelevant in the field theory limit.

It is now useful to introduce the variables
\beq
\nu_i = - \frac{1}{2} \log z_i ~~~~~ {\rm and}~~~~~
\tau = - \frac{1}{2} \log k~~,
\label{vari}
\eeq
so that \eq{oneloop1} becomes
\bea
A^{(1)}_M (p_1,\ldots,p_M) & = & N \, {\rm Tr}(\lambda^{a_1}
\cdots \lambda^{a_M}) ~
\frac{1}{(4\pi)^{d/2}}\left(\frac{g}{4}\right)^M
 ( 2 \alpha ')^{M-d/2} \nl
& \times & \int_{0}^{\infty}  d\tau ~
{\rm e}^{- 2 \a 'm^{2} \tau} \, \tau^{-d/2}
\prod_{n=1}^\infty \left(1 - {\rm e}^{-2 n \tau} \right)^{-d} \nl
& \times & \int_{0}^{\tau} d \nu_{M} \int_{0}^{\nu_{M}} d \nu_{M-1} \dots
\int_{0}^{\nu_3} d \nu_2 ~ \prod_{i=1}^{M}  \left[ \frac{z_i}{ V_{i} ' (0)}
\right]^{a - \a ' p_{i}^{2}}\label{oneloop2} \\
& \times & \prod_{i<j} \Big[\exp(G(\nu_{ji}))\Big]^{
2\a' p_i\cdot p_j}\, \prod_{i=2}^{M+1} \Big[ 1 -
 \exp(2(\nu_{i-1} -
 \nu_{i} ))\Big]^{a-1}~~,
\nonumber
\ena
where $G( \nu_{ji})$ is the one--loop bosonic Green function given in
Ref.~\cite{old1} with $\nu_{ji} \equiv \nu_j - \nu_i$, and
in the last product we have used the notation $\nu_{M+1} \equiv \tau$.
As in the case of the one--loop $M$--gluon amplitude, here too we find
a dependence on the local coordinates $V_{i} ' (0)$.
This can
be eliminated either by going on shell, or by choosing
$V_{i} ' (0) = z_i$ and formally staying off shell.
We will now show that \eq{oneloop2} can be used to obtain in a
simple and direct way the one--loop 1PI
diagrams of the $\Phi^3$ field theory defined by the
Lagrangian (\ref{Phi3lag}).

\sect{The field theory limit of scalar amplitudes}
\label{filiamp}

If we neglect the pinching configurations that yield the one-particle
reducible diagrams, as explained in great detail in Ref.~\cite{old1}, the
field theory limit of the one-loop
amplitude (\ref{oneloop2}) is simply
obtained by sending both $\nu_i$ and $\tau$ to $ \infty$.
This follows from the fact that, if we introduce the dimensionful variables
\beq
t = 2 \a' \, \tau  ~~~~~{\rm and}~~~~ x_i = 2 \a ' \, \nu_i~~,
\label{filim}
\eeq
the field theory is recovered in the limit $\a' \rightarrow 0$
keeping the ``proper times'' $t$ and $x_i$ finite.
In this case we can approximate the one-loop bosonic Green function
with~\cite{strass,scsc}
\beq
G (\nu) \rightarrow \nu - \frac{\nu^2}{\tau}~~,
\label{filimi}
\eeq
so that \eq{oneloop2} drastically simplifies.
In fact, observing that the
last terms in \eq{oneloop2} are negligible in
this limit, we obtain
\bea
A^{(1)}_M (p_1,\ldots,p_M) & = & N \, {\rm Tr}(\lambda^{a_1}
\cdots \lambda^{a_M}) \, \frac{1}{(4\pi)^{d/2}}
\left( \frac{g}{4}  \right)^{M}  \nl
& \times & \int_{0}^{\infty}  dt ~  {\rm e}^{- m^2 t} ~ t^{M-1-d/2} \nl
& \times & \int_{0}^{1} d {\hat{\nu}}_{M} \int_{0}^{{\hat{\nu}}_{M}} d
{\hat{\nu}}_{M-1} \cdots \int_{0}^{{\hat{\nu}}_3} d {\hat{\nu}}_2
\label{oneloopfith} \\
& \times & \prod_{i<j} \Bigg[
\exp \Big( p_i \cdot p_j \,t\, {\hat{\nu}}_{ji} ( 1- {\hat{\nu}}_{ji})
\Big)\Bigg]~~,
\nonumber
\ena
where, as in Ref.~\cite{old1}, we have chosen  $V_{i} ' (0) = z_i$
in order to reproduce the field theory limit also off mass--shell, and
defined ${\hat{\nu}}_{ji}={\nu}_{ji}/\tau$.

Both the integrand and the integration region in \eq{oneloopfith}
agree with the irreducible $\Phi^3$ field theory diagram
obtained from the Lagrangian (\ref{Phi3lag}), after the exponentiation
of each propagator of the diagram by means of the corresponding
Schwinger proper time
\beq
\frac{1}{ p^2 + m^2 } = \int_{0}^{\infty} dx ~e^{-x ( p^2 + m^2 )}~~,
\label{Schwi}
\eeq
and after the integration over the loop momentum.
The variable $t$ corresponds to the total Schwinger proper time of
the diagram, {\it i.e.} the sum of all Schwinger proper times,
while the variables ${\hat{\nu}}_{i}$ correspond to the
fraction of a partial proper time relative to the total one,
namely
\beq
t \equiv \sum_{i=1}^{M} t_i~~~~~,~~~~~ {\hat{\nu}}_{i} \equiv
\frac{t_2 + t_3 + \dots + t_i}{t}~~~~~{\rm for}~~~ i=2, \dots, M~~.
\label{tra}
\eeq
This shows that the field theory limit of the one--loop $M$--point
scalar amplitudes of the dual model correctly reproduces the
one--loop 1PI diagram with $M$ legs of the $\Phi^3$ field theory.

In principle, also multiloop amplitudes could be computed
in this pre-string dual model, using for instance the sewing
procedure.  However, since our main interest
here is to understand which regions in moduli space lead to the different
field theory diagrams, with the aim to later apply this analysis to
Yang-Mills theory, in this paper we will use a simpler ansatz, which is a
natural generalization of \eq{oneloop2}, and leads to the correct results.
In practice, we will perform the field theory limit directly in the amplitudes
(\ref{hmastac})
(corresponding to the case $a=1$), keeping only the lowest tachyonic state;
we then transform it into a normal scalar particle with arbitrary mass $m$
by rewriting each quadratic pole of measure (\ref{hmeasure})
according to
\beq
x^{-2} \rightarrow x^{-1-a}=x^{-1}\exp \left[-a \log x \right]=
x^{-1}\exp \left[m^2 \a' \log x \right] ~~.
\label{mass}
\eeq

Let us then turn back to \eq{hmastac}, and focus on the case $h=2$.
Since we want only the scalars to propagate in the
loops, it is enough to
compute the various quantities that appear in \eq{hmastac} for
$k_1 , k_2 \rightarrow 0$. In this limit, the determinant of the period
matrix can be approximated by
\beq
\det ( - i \tau_{\mu \nu} ) = \frac{1}{4 \pi^2} \left[ \log k_1
\log k_2 - \log^2 S  \right]~~,
\label{permat}
\eeq
while the two--loop bosonic Green
function becomes
\beqa
{\cal{G}}^{(2)} (z_1 , z_2) & = & \log ( z_1 - z_2 ) + \frac{1}{2}
\left[ \log k_1  \log k_2 -  \log^2 S \right]^{-1} \label{Green} \\
& \times & \left[ \log^2 T \log k_2 + \log^2 U \log k_1  - 2 \log T \log U
\log S \right]~~,   \nonumber
\eeqa
where we defined the anharmonic ratios
\beqa
S & = & \frac{(\eta_1 - \eta_2 ) (\xi_1 - \xi_2 )}{(\xi_1 - \eta_2 )
(\eta_1 - \xi_2 )}~~, \nonumber \\
T & = & \frac{(z_2 - \eta_1 ) (z_1 - \xi_1 )}{(z_2 - \xi_1 )
(z_1 - \eta_1 )}~~,  \label{STU}  \\
U & = &\frac{(z_2 - \eta_2 ) (z_1 - \xi_2 )}{(z_1 - \eta_2 )
(z_2 - \xi_2 )}~~. \nonumber
\eeqa
Notice finally that in the field theory limit of the scalar
amplitudes, the infinite
product over primary classes of the Schottky group is negligible and can be
simply replaced by 1.
We can thus express the two--loop $M$--point scalar amplitude as
\bea
A^{(2)}_M (p_1,\ldots,p_M) & = & N^2\,{\rm Tr}(\lambda^{a_1}
\cdots \lambda^{a_M})\,
\frac{1}{(4\pi)^d}\,\frac{g^{2+M}}{2^{8+3M}}\,
(2 \a' )^{3-d+M} \nl
& \times & \int \frac{ dk_1}{k_{1}^{2}} \int \frac{ d k_2}{k_{2}^{2}}
\int \frac{ d \xi_1 ~d \eta_1 }{ ( \xi_1 -  \eta_1 )^{2} }
\prod_{i=1}^{M} \left[ \frac{d z_i}{V_{i} ' (0) } \right]
\frac{1}{d \rho_{c}}  \label{2Mtac}\\
& \times &
\left[\frac{1}{4} \left(\log k_1 \log k_2 - \log^2 S \right)
\right]^{-d/2} \nl
& \times & \prod_{i<j} \left[{{\exp\left({\cal G}^{(2)}(z_i,z_j)\right)}
\over{\sqrt{V'_i(0)\,V'_j(0)}}}\right]^{2\a' p_i\cdot p_j}~~,
\nonumber
\ena
where ${\cal G}^{(2)}(z_i,z_j)$ is given by \eq{Green}, with
$\xi_2= \infty$ and $ \eta_2 =0$.
For future convenience, we have left undetermined
the variable $\rho_c$, that eventually will be
fixed to $1$.
This formula is the master formula that can be used to calculate
all two--loop diagrams in the $\Phi^3$
field theory, as we will now show.

Let us start from the two--loop vacuum bubbles ($M=0$).
In this case, choosing $\rho_c=\xi_1=1$,  \eq{2Mtac}
simply becomes \footnote{The overall $N^3$ factor comes from
$N^2\,{\rm Tr}(1)$.}
\bea
A^{(2)}_0 & = & \frac{N^3}{(4\pi)^d}\,\frac{g^{2}}{2^{8}}\,
(2 \a' )^{3-d}
\int \frac{dk_1}{k_{1}^{2}} \int \frac{ d k_2}{k_{2}^{2}}
\int \frac{d \eta_1 }{ (1 -  \eta_1 )^{2}} \nonumber \\
& \times & \left[\frac{1}{4} \left(\log k_1 \log k_2 -
 \log^2 \eta_1 \right) \right]^{-d/2} \label{vacuum}~~.
\ena
As expected, the two multipliers $k_1$ and $k_2$
play the same role, since
\eq{vacuum} is symmetrical in the exchange of $k_1$ and $k_2$.
Therefore, to avoid double counting, we order
them by chosing, for example, $k_2 \leq k_1$.
The integration region of the third modular parameter, $\eta_1$, can be deduced
by studying in more detail the Schottky representation of the two--annulus,
which
is shown in Fig. 1.
The points $A$, $B$, $C$ and $D$ have to be identified with
$A'$, $B'$, $C'$ and $D'$ respectively, under the action of the two
generators of the Schottky group.

\vspace{1.5cm}
\unitlength 1mm
\linethickness{0.4pt}
\begin{picture}(125.00,40.00)(10,75)
\put(80.00,100.00){\circle{8.67}}
\put(100.00,100.00){\circle{12.00}}
\put(126.00,100.00){\circle{12.00}}
\multiput(150.00,120.00)(0.11,-0.61){4}{\line(0,-1){0.61}}
\multiput(150.46,117.56)(0.10,-0.61){4}{\line(0,-1){0.61}}
\multiput(150.86,115.12)(0.11,-0.81){3}{\line(0,-1){0.81}}
\multiput(151.20,112.68)(0.09,-0.81){3}{\line(0,-1){0.81}}
\multiput(151.48,110.24)(0.11,-1.22){2}{\line(0,-1){1.22}}
\multiput(151.70,107.80)(0.08,-1.22){2}{\line(0,-1){1.22}}
\put(151.86,105.37){\line(0,-1){2.44}}
\put(151.96,102.93){\line(0,-1){2.44}}
\put(152.00,100.49){\line(0,-1){2.44}}
\put(151.98,98.05){\line(0,-1){2.44}}
\multiput(151.90,95.61)(-0.07,-1.22){2}{\line(0,-1){1.22}}
\multiput(151.77,93.17)(-0.10,-1.22){2}{\line(0,-1){1.22}}
\multiput(151.57,90.73)(-0.09,-0.81){3}{\line(0,-1){0.81}}
\multiput(151.31,88.29)(-0.11,-0.81){3}{\line(0,-1){0.81}}
\multiput(151.00,85.85)(-0.09,-0.61){4}{\line(0,-1){0.61}}
\multiput(150.62,83.41)(-0.10,-0.57){6}{\line(0,-1){0.57}}
\put(155.00,100.00){\line(-1,0){150.00}}
\multiput(10.00,80.00)(-0.11,0.61){4}{\line(0,1){0.61}}
\multiput(9.58,82.44)(-0.09,0.61){4}{\line(0,1){0.61}}
\multiput(9.21,84.88)(-0.11,0.81){3}{\line(0,1){0.81}}
\multiput(8.89,87.32)(-0.09,0.81){3}{\line(0,1){0.81}}
\multiput(8.63,89.76)(-0.11,1.22){2}{\line(0,1){1.22}}
\multiput(8.41,92.20)(-0.08,1.22){2}{\line(0,1){1.22}}
\put(8.25,94.63){\line(0,1){2.44}}
\put(8.15,97.07){\line(0,1){2.44}}
\put(8.09,99.51){\line(0,1){2.44}}
\put(8.08,101.95){\line(0,1){2.44}}
\put(8.13,104.39){\line(0,1){2.44}}
\multiput(8.23,106.83)(0.08,1.22){2}{\line(0,1){1.22}}
\multiput(8.38,109.27)(0.10,1.22){2}{\line(0,1){1.22}}
\multiput(8.59,111.71)(0.09,0.81){3}{\line(0,1){0.81}}
\multiput(8.84,114.15)(0.10,0.81){3}{\line(0,1){0.81}}
\multiput(9.15,116.59)(0.10,0.68){5}{\line(0,1){0.68}}
\put(149.17,102.00){\makebox(0,0)[cc]{$B'$}}
\put(11.50,102.00){\makebox(0,0)[cc]{$A'$}}
\put(73.60,102.00){\makebox(0,0)[cc]{$A$}}
\put(86.20,102.00){\makebox(0,0)[cc]{$B$}}
\put(92.00,102.00){\makebox(0,0)[cc]{$C$}}
\put(108.67,102.00){\makebox(0,0)[cc]{$D$}}
\put(117.50,102.00){\makebox(0,0)[cc]{$D'$}}
\put(135.20,102.00){\makebox(0,0)[cc]{$C'$}}
\put(80,108.00){\makebox(0,0)[cc]{${\cal K}_2$}}
\put(100.70,110.00){\makebox(0,0)[cc]{${\cal K}_1$}}
\put(126.70,110.00){\makebox(0,0)[cc]{${\cal K}_1'$}}
\end{picture}

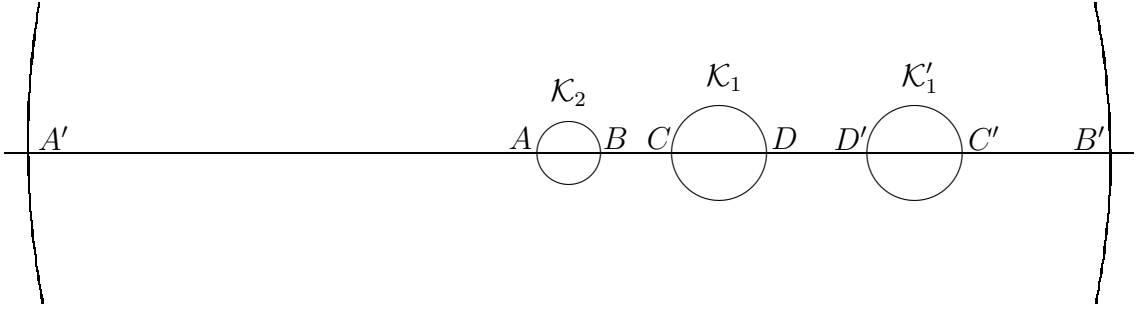
Fig. 1: In the Schottky parametrization, the two--annulus corresponds to the
part of the upper--half plane which is inside the big
circle passing through $A'$ and $B'$, and which is outside the
circles ${\cal K}_1$, ${\cal K}_1'$ and ${\cal K}_2$.
\vspace{1.0cm}

The position of these points is completely determined once
$k_1$, $k_2$ and $\eta_1$ are given; for example,
following \cite{scho}, one can verify that
\beq
B=\sqrt{k_2}~~,~~~ C={\eta_1-\sqrt{k_1}\over
1-\sqrt{k_1}}~~,~~~D' -D={1-\sqrt{k_1}\over
1+\sqrt{k_1}}\,(1-\eta_1)~~.
\eeq
It is easy to realize that
the two segments $(AA')$ and $(DD')$ represent
respectively the two inner boundaries of the
two--annulus, while the union of
$(BC)$ and $(C'B')$ represents the external boundary.
With these choices, the interpretation of the three moduli,
$k_1$, $k_2$ and $\eta_1$, is particularly simple. In fact,
$\sqrt{k_2}$ is the radius of the circle ${\cal K}_2$, while the radii of
${\cal K}_1$ and ${\cal K}_1'$
are proportional to $\sqrt{k_1}$. Furthermore,
$\eta_1$ turns out to be inside ${\cal K}_1$, while
the point $\xi_1=1$ is inside ${\cal K}_1'$.
Therefore, in this configuration the  circle ${\cal K}_1'$ is fixed while
${\cal K}_1$ can move, depending on the value of $\eta_1$.
In particular, if the point $D'$ is very
close to $D$, $\eta_1$ is almost equal to
1, while if $C$ is near to $B$, then $\eta_1$ is slightly
bigger than $\sqrt{k_1}\to 0$.
Note that in this way we have just found for $k_1$, $k_2$ and $\eta_1$
the same integration region of integration derived by
Roland \cite{kajpr} for the closed string,
{\it i.e.} $0\leq\sqrt{k_2}\leq \sqrt{k_1}\leq\eta_1\leq 1$.
These facts allow us to interpret $\eta_1$
as the ``distance'' between the two loops. Thus,
when $\eta_1\to 1$ we expect to eventually find from \eq{vacuum}
the reducible vacuum bubble of the $\Phi^3$ field theory with
the two loops widely separeted, while when $\eta_1\to \sqrt{k_1}\to 0$
we expect to obtain from \eq{vacuum} the irreducible vacuum bubble with
the two loops attached to each other. In what follows we will show that
this is indeed what happens.

In fact, when $\eta_1\rightarrow 1$ (together with $k_1,k_2\to 0$),
we have
\bea
A^{(2)}_0\Big|_{\rm red} & = &
\frac{N^3}{(4\pi)^d}\,\frac{g^{2}}{2^{8}}\,
(2 \a' )^{3-d}
\!\int_{1-\epsilon}^{1} \frac{d \eta_1 }{ (1 -  \eta_1)^{2}}
\int_0^{\epsilon} \frac{dk_2}{k_{2}^{2}}\int_{k_2}^{\epsilon} \frac{ d
k_1}{k_{1}^{2}}  \left[\frac{1}{4}\left(\log k_1 \log k_2 \right)
\right]^{-d/2}\nonumber\\
& = &  \frac{N^3}{(4\pi)^d}\,\frac{g^{2}}{2^{5}}\,
\int_0^{\infty}dt_3 \int_0^{\infty}dt_2 \int_0^{t_2}dt_1~
{\rm e}^{-m^2(t_1+t_2+t_3)}
{}~(t_1\,t_2)^{-d/2}~~,\label{vacuumr}
\ena
where in the second line we have introduced the mass $m$
as explained in \eq{mass}, and the Schwinger proper times $t_i$
according to
\beq
t_1=-\a'\log k_1\,,~~~t_2=-\a'\log k_2\,,~~~t_3=-\a'\log (1-\eta_1)~~.
\label{ti}
\eeq
Since \eq{vacuumr} is symmetrical in $t_1$ and $t_2$, it is possible to perform
the
integration over $t_1$ and $t_2$ independently from 0 to $\infty$
by introducing a
factor of $1/2$. In this way one obtains exactly the same result of the
reducible vacuum bubble of the $\Phi^3$ field theory defined by
\eq{Phi3lag}.

In the second case, $\eta_1\to 0$, it is more convenient to introduce
\beq
q_1={k_2\over\eta_1}~,~~~q_2={k_1\over\eta_1}~,~~~q_3=\eta_1~~,
\label{qvariable}
\eeq
so that \eq{vacuum} becomes
\bea
A^{(2)}_0\Big|_{\rm irr} & = &  \frac{N^3}{(4\pi)^d}\,\frac{g^{2}}{2^{8}}\,
(2 \a' )^{3-d}
\!
\int_0^{\epsilon} \frac{dq_3}{q_{3}^{2}} \int_{0}^{q_3} \frac{ d
q_2}{q_{2}^{2}} \int_{0}^{q_2}\frac{ d q_1}{q_{1}^{2}}
\nonumber\\
&\times&\left[\frac{1}{4}\left(\log q_1 \log q_2 +\log q_1 \log q_3
+\log q_2 \log q_3 \right)\right]^{-d/2}\label{vacuumir} \\
& = &  \frac{N^3}{(4\pi)^d}\,\frac{g^{2}}{2^{5}}
\!
\int_0^{\infty}dt_3 \int_0^{t_3}dt_2 \int_0^{t_2}dt_1~
e^{-m^2(t_1+t_2+t_3)}\nl
&\times& (t_1t_2+t_1t_3+t_2t_3)^{-d/2}~~.
\nonumber
\ena
Here, again,  we have introduced the Schwinger proper times $t_i$ related to
$q_i$ variables as in \eq{ti}.
Since \eq{vacuumir} is completely symmetrical, we can introduce a factor of
$1/3!$ and perform the three integrals
independently from 0 to $\infty$. In this way we
correctly
reproduce the irreducible vacuum bubble of the $\Phi^3$ theory.
Note that by using a single starting
formula, namely \eq{vacuum}, we have been able
to obtain two diagrams which have a different weight;
this has been
possible because \eq{vacuum} has different symmetry properties in
the two regions of the moduli space that yield
the two vacuum bubbles of the $\Phi^3$
field theory.

We now turn to the one--point amplitude $(M=1)$. In this case,
momentum conservation prevents us from imposing the mass shell condition, so
that the final result will depend on the choice of the local
coordinates $V_{1} ' (0)$ around the puncture $z_1$. This issue requires a
thorough discussion which we leave to a future publication. Here we simply
propose a  generalization of the one--loop  choice ($V_{i} ' (0) = z_i$), that
is strongly inspired by
Ref.~\cite{kajsat} and that leads to consistent results.
We propose to define
\beq
{\left( V_{i} ' (0) \right)}^{-1} = \left\vert{1\over z_i - \rho_a} -
{1\over z_i - \rho_b}\right\vert~~, \label{Vi}
\eeq
where $\rho_a$ and $\rho_b$ depend on the position of $z_i$ and are
the two fixed points that stay on the left and on the right of $z_i$. For
example, referring to Fig. 1, if $z_i$ is between
$D$ and $D'$ we have $\rho_a=\eta_1$ and $\rho_b=\xi_1=1$, while
if $z_i$ is between $C'$ and $B'$
we have $\rho_a=\xi_1=1$ and $\rho_b=\xi_2=\infty$.
Note that at one loop in the standard configuration
we have $\rho_a=\eta=0$ and
$\rho_b=\xi=\infty$, and thus, in this case, \eq{Vi} reduces to
$V'_i(0)=z_i$.
We remark that the local coordinates in \eq{Vi} are chosen
at the string level {\it before} taking the field
theory limit, and as such, they have a general validity. However,
since different regions of the moduli space ({\it i.e.} different values
of $\rho_a$ and $\rho_b$) usually correspond to different Feynman diagrams in
the field theory limit, the explicit expression of $V'_i(0)$ may look
different for different diagrams, precisely like in the analysis
of Ref.~\cite{kajsat}.

Using the local coordinates (\ref{Vi}), and the integration region of
the moduli space determined from the study of the vacuum bubbles, we can
compute the two--loop one--point amplitude explicitly
starting from \eq{2Mtac} with $M=1$ and
$\rho_c=\xi_1=1$. The
result of this calculation, which we will describe in
detail elsewhere, is
\bea
A^{(2)}_1 & = & -N^2\,{\rm Tr}(\lambda^{a_1})\,
\frac{1}{(4\pi)^d}\,\frac{g^{3}}{2^{10}}\,
(2 \a' )^{4-d}
\!
\int_0^{1} \frac{d q_3 }{ (1 - q_3)^{2}~ q_3^2}
 \int_{0}^{q_3}\frac{dq_2}{q_{2}^{2}} \int_{0}^{q_2}
\frac{d q_1 }{ q_1^2}
\nonumber\\ & \times & \left[\frac{1}{4}\left(\log q_1 \log
q_2 + \log q_1 \log q_3 + \log q_2 \log q_3 \right)
\right]^{-d/2}\nonumber\\ \label{1point}
& \times & \Big[\log q_1 + \log q_2 + \log q_3+ \log (1-q_3)\Big]~~,
\ena
where $q_1$, $q_2$ and $q_3$ are defined in \eq{qvariable}.
{}From \eq{1point} one can derive the irreducible two--loop one--point
function of the $\Phi^3$ field
theory when $q_3\to q_2\to q_1\to 0$, while when $q_3 \rightarrow 1$ and
$q_2\to q_1\to 0$ one obtains the
two reducible diagrams.

We have explicitly verified that this method gives the correct results
also when there are two external states. However,
since in this case there are many corners of the string moduli space
that contribute to the
same field theory diagram, the analysis is a bit lengthy,
and so we leave it to a forthcoming publication.
Here instead, we propose a different procedure inspired by
Ref.~\cite{Fabbri} that also leads to the correct identification of the
field theory diagrams starting from \eq{2Mtac}.

Let us start from the two-loop diagram depicted in Fig. 2.

\vspace{1cm}

\unitlength 1.2mm
\linethickness{0.4pt}
\begin{picture}(104.67,16.00)(20,110)
\put(70.00,120.00){\circle{12.00}}
\put(90.00,120.00){\circle{12.00}}
\put(76.00,120.00){\line(1,0){8.00}}
\put(96.00,120.00){\line(1,0){8.67}}
\put(64.00,120.00){\line(-1,0){9.00}}
\end{picture}

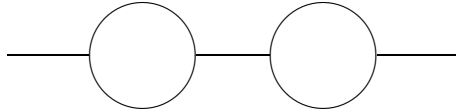
Fig. 2: A reducible two--loop diagram in the $\Phi^3$
field theory contributing to the two--point function.
\vspace{1cm}

First we cut open the two loops of the
diagram, and subdivide the resulting tree diagram so that it can be
obtained by sewing three-point vertices. Next we fix the legs of each
three--point vertex at $0,1$ and $\infty$, as depicted in Fig. 3.

\vspace{1cm}

\unitlength .75mm
\linethickness{0.4pt}
\begin{picture}(50.00,100.00)(-23,100)
\put(16.67,186.67){\line(0,-1){15.00}}
\put(16.67,186.67){\line(1,0){15.00}}
\put(16.67,186.67){\line(0,1){15.00}}
\put(16.67,204.67){\makebox(0,0)[cc]{$\xi_2=\infty$}}
\put(40.17,186.67){\makebox(0,0)[cc]{$z_1=1$}}
\put(16.67,168.67){\makebox(0,0)[cc]{$0$}}
\put(19.67,159.67){\makebox(0,0)[cc]{$A_1$}}
\put(16.67,131.67){\line(0,-1){15.00}}
\put(16.67,131.67){\line(0,1){15.00}}
\put(16.67,131.67){\line(1,0){15.00}}
\put(16.67,149.67){\makebox(0,0)[cc]{$\infty$}}
\put(34.17,131.67){\makebox(0,0)[cc]{$1$}}
\put(16.67,113.67){\makebox(0,0)[cc]{$\eta_2=0$}}
\put(40.67,126.67){\makebox(0,0)[cc]{$A_2$}}
\put(65.67,131.67){\line(-1,0){15.00}}
\put(65.67,131.67){\line(1,0){15.00}}
\put(65.67,131.67){\line(0,1){15.00}}
\put(47.17,131.67){\makebox(0,0)[cc]{$\infty$}}
\put(65.67,149.67){\makebox(0,0)[cc]{$\eta_1=0$}}
\put(83.17,131.67){\makebox(0,0)[cc]{$1$}}
\put(89.17,126.67){\makebox(0,0)[cc]{$A_3$}}
\put(114.67,131.67){\line(-1,0){15.00}}
\put(114.67,131.67){\line(1,0){15.00}}
\put(114.67,131.67){\line(0,-1){15.00}}
\put(96.17,131.67){\makebox(0,0)[cc]{$\infty$}}
\put(138.67,131.67){\makebox(0,0)[cc]{$\xi_1=1$}}
\put(114.67,113.67){\makebox(0,0)[cc]{$z_2=0$}}
\end{picture}

Fig. 3: The sewing configuration of three-string vertices
corresponding to the diagram of Fig. 2.
\vspace{1cm}

Then, we reconnect the diagram by inserting between vertices a
suitable propagator acting on the external legs as a well--specified
projective transformation. This transformation is chosen in such a way
that its fixed points are
the Koba-Nielsen variables of the two legs that are sewn.
For example, if we sew two legs
corresponding to the points $0$ and $\infty$,
we use the transformation
\beq
S(z) = A z~~,
\label{projtra1}
\eeq
whereas, if we sew two legs corresponding to $1$ and $\infty$, we
use
\beq
S(z) = A z + 1 - A~~.
\label{projtra2}
\eeq
Finally, we sew together the legs corresponding to $\xi_1$ and
$\eta_1$, and to $\xi_2$ and $\eta_2$ by means of two projective
transformations
with multipliers $k_1$ and $k_2$ respectively.
In this way we recover the full two--loop
diagram.

After the sewing has been performed, both the Koba-Nielsen variables
and the Schottky fixed points become functions of the parameters $A_i$
that appear in the various projective transformations.
For example, for the particular sewing configuration
depicted in Fig. 3, we obtain
\beq
\xi_1 = A_1~~, \hspace{0.8cm} z_2 = A_1 (1 - A_2 A_3)~~, \hspace{0.8cm}
\eta_1 = A_1 (1 - A_2)~~, \hspace{0.8cm} z_1 = 1~~,
\label{newvar1}
\eeq
together with $\xi_2 = \infty$ and $\eta_2 =0$.

The parameters $A_i$ have a simple geometric interpretation, and
drive the field theory limit. In fact,
as described in Ref.~\cite{Fabbri}, they are related to
the length of the strip connecting
two three--point vertices. In the
limit $\alpha' \to 0$, this length
is given by
\beq
t_i = - \a ' \log A_i~~,  \hspace{2cm} i=1,2,3~~,
\label{ai}
\eeq
while the lengths of the two entire loops are similarly related
to the multipliers $k_1$ and $k_2$.
The field theory limit is thus the limit $A_i \to 0$ and
$k_\mu \to 0$, with
a definite ordering prescribed by the sewing procedure. Notice that,
although the result of this procedure does not depend on how
a given diagram is cut, for the explicit calculations
it is always convenient to choose a particular
sewing configuration, like for example the one of Fig. 3
corresponding to the diagram of Fig. 2.
Then, the limit $A_i \ll 1$ implies that
the original variables of the
diagram in Fig. 2 must be ordered as follows
\beq
\xi_2= \infty \gg z_1 =1 \gg \xi_1 \gg z_2 \gg \eta_1 \gg \eta_2 = 0~~.
\label{pin1}
\eeq
The proper times associated to individual propagators are given by
\eq{ai}, and by
\beq
t_4 = - \a ' \log \frac{k_2}{A_1}~~,  \hspace{2cm}
t_5 = - \a ' \log \frac{k_1}{A_3}~~.
\label{k121}
\eeq
We now use \eq{mass} to eliminate the double poles
in the measure of integration which, in terms of the proper times
$t_i$, becomes
\beq
[dm]^{2}_{2} \rightarrow  - 2^5 (2 \a ')^{d -5} \prod_{i=1}^{5} d
t_i\,
\Big[ (t_3 + t_5 ) ( t_1 + t_4) \Big]^{-d/2}
\, e^{- m^2 (t_5 + t_4)} \Big[ V_{1} ' (0) V_{2} ' (0)
\Big]^{\a' m^2}~~.
\label{mea1}
\eeq
On the other hand, in the region \eq{pin1}, the Green
function ${\cal{G}}^{(2)}$ simplifies according to
\beq
{\cal{G}}^{(2)} \rightarrow  - \frac{1}{2 \a '}
\left[ \frac{t_{1}^{2}}{ t_1 + t_4} + \frac{t_{3}^{2}}{ t_3 + t_5} \right]~~.
\label{gree21}
\eeq
If we insert Eqs. (\ref{mea1}) and (\ref{gree21})
into \eq{2Mtac}, with
$M=2$ and $\rho_c=z_1=1$, we see that
all factors of $\a'$ cancel, so that we are left with
\bea
A^{(2)}_2 (p_1, p_2) & \rightarrow & N^2\,{\rm Tr}(\lambda^{a_1}
\lambda^{a_2})\,\frac{1}{( 4 \pi)^d}
\,\frac{g^{4}}{2^{9}} \int_{0}^{\infty} d t_2 ~e^{- t_2 ( p^2 + m^2)}  \nl
& \times & \prod_{i \neq 2}
\int_{0}^{\infty} d t_i ~e^{- m^2 ( t_1 + t_3 + t_4 + t_5 )}
\left[ (t_3 + t_5 ) ( t_1 + t_4) \right]^{-d/2}  \label{2tac1}\\
& \times & \exp\Bigg\{
- p^2 \left[ \frac{t_1 t_4}{t_1 + t_4} +
\frac{t_3 t_5}{t_3 + t_5} \right]\Bigg\}\nl
&\times&\exp\Bigg\{(p^2 + m^2 ) \Big[ t_1 + t_2 + t_3 + \a ' \log \left(
V_{1} ' (0) V_{2} ' (0) \right) \Big]\Bigg\}~~,
\nonumber
\ena
where, by momentum conservation, $p_1 = - p_2 \equiv p$.

One can easily check that \eq{2tac1} correctly reproduces
the field theory diagram in Fig. 2, including the overall factor.
In particular, if we put
the external legs on the mass shell, the dependence on
$V_{i} ' (0)$ disappears like in the full string amplitude.
On the other hand, if we use the
coordinates  $V_{i} ' (0)$ of \eq{Vi}, we can reproduce
the correct field theory
result also off the mass--shell.

Let us consider now the 1PI diagram in Fig. 4.

\vspace{1cm}

\unitlength 1.2mm
\linethickness{0.4pt}
\begin{picture}(95.33,15.00)(15,110)
\put(80.00,120.00){\circle{12.00}}
\put(80.00,126.00){\line(0,-1){12.00}}
\put(86.00,120.00){\line(1,0){9.33}}
\put(74.00,120.00){\line(-1,0){9.00}}
\end{picture}

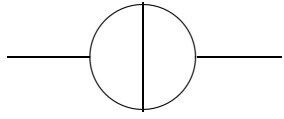
Fig. 4: An irreducible two-loop diagram contributing to the two--point
function of the $\Phi^3$ theory.

\vspace{1cm}

Applying our procedure, we first open
the two loops, and then we sew the corresponding tree diagrams in the
configuration shown in Fig. 5.

\vspace{1cm}

\unitlength 0.75mm
\linethickness{0.4pt}
\begin{picture}(80.00,80.00)(-25,115)
\put(16.67,179.67){\line(0,-1){15.00}}
\put(16.67,179.67){\line(1,0){15.00}}
\put(16.67,179.67){\line(0,1){15.00}}
\put(16.67,197.67){\makebox(0,0)[cc]{$\xi_1=1$}}
\put(40.17,179.67){\makebox(0,0)[cc]{$z_1=0$}}
\put(16.67,162.27){\makebox(0,0)[cc]{$\infty$}}
\put(21.67,155.67){\makebox(0,0)[cc]{$A_1$}}
\put(16.67,131.67){\line(-1,0){15.00}}
\put(16.67,131.67){\line(0,1){15.00}}
\put(16.67,131.67){\line(1,0){15.00}}
\put(16.67,149.87){\makebox(0,0)[cc]{$1$}}
\put(34.17,131.67){\makebox(0,0)[cc]{$0$}}
\put(-7.67,131.67){\makebox(0,0)[cc]{$\xi_2=\infty$}}
\put(40.67,126.67){\makebox(0,0)[cc]{$A_2$}}
\put(64.67,131.67){\line(-1,0){15.00}}
\put(64.67,131.67){\line(1,0){15.00}}
\put(64.67,131.67){\line(0,1){15.00}}
\put(45.67,131.67){\makebox(0,0)[cc]{$\infty$}}
\put(64.67,149.67){\makebox(0,0)[cc]{$z_2=1$}}
\put(82.17,131.67){\makebox(0,0)[cc]{$0$}}
\put(88.67,126.67){\makebox(0,0)[cc]{$A_3$}}
\put(112.67,131.67){\line(-1,0){15.00}}
\put(112.67,131.67){\line(1,0){15.00}}
\put(112.67,131.67){\line(0,1){15.00}}
\put(93.67,131.67){\makebox(0,0)[cc]{$\infty$}}
\put(137.67,131.67){\makebox(0,0)[cc]{$\eta_2=0$}}
\put(112.67,149.67){\makebox(0,0)[cc]{$\eta_1=1$}}
\end{picture}

Fig. 5: The sewing configuration of three-string vertices
corresponding to the diagram of Fig. 4.
\vspace{1cm}

Following the same steps as in the previous case,
we obtain
\beq
z_1 = 1 - A_1~~, \hspace{0.8cm} z_2 = A_2~~, \hspace{0.8cm}
\eta_1 = A_2 A_3~~, \hspace{0.8cm} \xi_1 = 1~~,
\label{newvar2}
\eeq
together with $\eta_2 =0$ and $\xi_2 =\infty$.
Since the variables $A_i$ must be constrained to be
very close to $0$, the variables of the
diagram of Fig. 4 are ordered as follows
\beq
\xi_2 = \infty \gg \xi_1 =1 \gg z_1 \gg z_2 \gg \eta_1 \gg \eta_2 = 0~~.
\label{pin2}
\eeq
The proper times are given again by \eq{ai}, together with
\beq
t_4 = - \a' \log \frac{k_2}{A_2 A_3}~~, \hspace{2cm}
t_5 = - \a' \log \frac{k_1}{A_1 A_2 A_3}~~.
\label{k122}
\eeq
In terms of these variables, the measure becomes
\bea
[dm]^{2}_{2} & \rightarrow & 2^5 (2 \a ' )^{d -5} \prod_{i=1}^{5} d t_i~
e^{- m^2 ( t_3 + t_4 + t_5 )} \prod_{i=1}^{2} e^{\a' m^2 \log V_{i} ' (0) }
\nl
& \times & \left[( t_2 + t_3 + t_4 ) (t_1 + t_2 + t_3 + t_5) -
( t_2 + t_3)^2
\right]^{-d/2}~~,
\label{mea2}
\ena
while the two--loop Green function in the region (\ref{pin2})  simplifies
according to
\beqa
& & {\cal{G}}^{(2)} \rightarrow \frac{1}{2 \a'}
\frac{1}{( t_2 + t_3 + t_4 ) (t_1 + t_2 + t_3 + t_5) -
( t_2 + t_3)^2}  \label{gree22}  \\
& \times & \left[ 2 (t_1 + t_2) ( t_2 + t_3) t_2
- ( t_2 + t_3 + t_4) ( t_1 + t_2 )^2 - t_{2}^{2}
(t_1 + t_2 + t_3 + t_5) \right]~~. \nonumber
\eeqa
Therefore, inserting Eqs.~(\ref{mea2}) and (\ref{gree22}) into
\eq{2Mtac}, with
$M=2$ and $\rho_c=\xi_1=1$, we get
\bea
&&A^{(2)}_2 (p_1, p_2) ~~\rightarrow ~~ N^2\,{\rm Tr}(\lambda^{a_1}
\lambda^{a_2})\, \frac{1}{( 4 \pi)^d }
\,\frac{ g^{4}}{ 2^{9}}  \int_{0}^{\infty} \prod_{i=1}^{5} d t_i~
e^{- m^2 \sum\limits_{i=1}^{5} t_i}  \nl
& \times &
\!\!\!\left[ ( t_2 + t_3 + t_4 ) (t_1 + t_2 + t_3 + t_5) - ( t_2 + t_3)^2
\right]^{-d/2}  \label{2tac2}\\
& \times & \!\!\!\exp\left\{
- p^2 \left[ \frac{ 2 (t_1 + t_2) ( t_2 + t_3) t_2
- ( t_2 + t_3 + t_4) ( t_1 + t_2 )^2 - t_{2}^{2}
(t_1 + t_2 + t_3 + t_5)}{( t_2 + t_3 + t_4 ) (t_1 + t_2 + t_3 + t_5) -
( t_2 + t_3)^2}   \right. \right.\nl
&&+ ~(t_1 + t_2) \Bigg] +
(p^2 + m^2 ) \Big[ t_1 + t_2  + \a ' \log \left(
V_{1} ' (0) V_{2} ' (0) \right) \Big]\Bigg\}~~.
\nonumber
\ena
Once again, the field theory diagram in Fig. 4 is correctly reproduced,
as one can easily check. If the external legs are kept on the
mass--shell, the dependence on $V_{i} ' (0)$ drops out;
on the other hand, if they are kept off
shell, the correct result is reproduced by choosing again
the $V_{i} ' (0)$'s of \eq{Vi}. We have also checked that this procedure allows
us to recover correctly all other two--loop
diagrams with two external legs.

The analysis presented in this paper shows that the open bosonic
string can be used to compute explicitly the amplitudes
of the $\Phi^3$ field theory including their overall normalization. These
results
are suited for several generalizations,
for example by increasing the number of external legs, or by increasing the
number of loops.
Furthermore, if, in the zero--slope limit, we select
the contributions of the vectors
instead of those of the scalars, our methods should allow to recover explicitly
the multiloop amplitudes of Yang--Mills theory from the open bosonic string.

\vskip 1.0cm

{\large {\bf {Acknowledgements}}}
\vskip 0.5cm
One of us (R. M.) would like to thank NORDITA for the kind hospitality.
This research was partially supported by MURST and the EU, within the framework
of the program ``Gauge Theories, Applied Supersymmetry and Quantum Gravity''
under contract SCI-CT92-0789.

\end{document}